\documentclass[conference]{IEEEtran}
\IEEEoverridecommandlockouts
\usepackage{cite}
\usepackage{amsmath,amssymb,amsfonts}
\usepackage{gensymb}
\usepackage{algorithmic}
\usepackage{graphicx}
\usepackage{textcomp}
\usepackage{xcolor}
\usepackage{subcaption}
\usepackage{url}
\usepackage{tikz}
\newcommand\copyrighttext{%
  \footnotesize \textcopyright 2024 IEEE. Personal use of this material is permitted. 
  Permission from IEEE must be obtained for all other uses, in any current or future 
  media, including reprinting/republishing this material for advertising or promotional 
  purposes, creating new collective works, for resale or redistribution to servers or 
  lists, or reuse of any copyrighted component of this work in other works.}

\newcommand\copyrightnotice{%
\begin{tikzpicture}[remember picture,overlay]
\node[anchor=south,yshift=10pt] at (current page.south) 
{\fbox{\parbox{\dimexpr\textwidth-\fboxsep-\fboxrule\relax}{\copyrighttext}}};
\end{tikzpicture}%
}

\usepackage{pgfplots}
\pgfplotsset{compat=1.17}
\usetikzlibrary{shapes.geometric, arrows}
\def\BibTeX{{\rm B\kern-.05em{\sc i\kern-.025em b}\kern-.08em
    T\kern-.1667em\lower.7ex\hbox{E}\kern-.125emX}}
\begin{document}

\title{RadarCNN: Learning-based Indoor Object Classification from IQ Imaging Radar Data\\

\thanks{The authors acknowledge the financial support by the Federal Ministry of Education and Research of Germany in the programme of “Souverän. Digital. Vernetzt.”. Joint project 6G-life, project identification number: 16KISK002}
}

\author{\IEEEauthorblockN{Stefan Hägele, Fabian Seguel, Driton Salihu, Marsil Zakour and Eckehard Steinbach}
\IEEEauthorblockA{\textit{Technical University of Munich} \\
\textit{School of Computation, Information and Technology}\\
\textit{Chair of Media Technology}\\
\textit{Munich Institute of Robotics and Machine Intelligence}\\
\{stefan.haegele, fabian.seguel, driton.salihu, marsil.zakour, eckehard.steinbach\}@tum.de}
}

\maketitle
\copyrightnotice

\begin{abstract}

Radar sensors operating in the mmWave frequency range face challenges when used as indoor perception and imaging devices, primarily due to noise and multipath signal distortions. These distortions often impair the sensors' ability to accurately perceive and image the indoor environment.
Nevertheless, this sensor offers distinct advantages over camera and LiDAR sensors.
This encompasses the estimation of object reflectivity, known as radar cross-section (RCS), and the ability to penetrate through objects that are thin or have low reflectivity.
This results in a 'through-the-wall' sensing capability.
Due to the aforementioned disadvantages, most research in the field of imaging radar tends to exclude indoor areas.
We introduce a machine learning-based mmWave MIMO FMCW imaging radar object classifier designed to identify small, hand-sized objects in indoor settings, utilizing only radar IQ samples as input. This system achieves 97-99\% accuracy on our test set and maintains approximately 50\% accuracy even under challenging conditions, such as increased background noise and occlusion of sample objects, without the need for adjusting training or pre-processing.
This demonstrates the robustness of our approach and offers insights into what needs to be improved in the future to achieve generalization and very high accuracy even in the presence of significant indoor perturbations.
\end{abstract}

\begin{IEEEkeywords}
mmWave radar, imaging radar, classification, signal processing, machine learning, deep learning, indoor environments, digital twin
\end{IEEEkeywords}

\section{Introduction}

The application of radar technology in several research areas has seen a significant increase in recent years. This surge can be attributed to the widespread availability of standard radar sensors and their relatively low purchase costs.
In general, research in the field of mmWave radar applications is currently focused on the automotive sector. Hence, radar systems are often used as a supplementary sensor for camera-based perception in vehicles. This approach is adopted to improve depth estimation and enhance visibility in adverse conditions \cite{mengchen1}, as well as to gain additional advantages by acquiring more comprehensive information about the environment through this sensor fusion method \cite{fusion1, fusion2}.
However, research focused on the perception of indoor spaces using mmWave radar lags behind that in the automotive sector. This is mainly due to the existence of several adverse indoor effects, notably strong multi-path reflections, which often result in ghost targets. In addition, the overall resolution is generally worse compared to LiDAR sensors or camera sensors \cite{lidarradar1, cameraradar1}.
However, the aforementioned advantages of radar, along with the added capability to estimate the RF reflectivity of various object surfaces and materials, makes the use of radar systems also appealing for indoor scenarios.
\\
We propose a new radar-based deep learning classification approach for small objects in indoor environments and evaluate its performance under three conditions. This includes classic testing as well as testing under impaired conditions like object occlusion and increased background noise. To simulate realistic conditions, we train our model on non-impaired data and test it on data under adverse and noisy conditions.
In this way, we demonstrate the robustness of the current system in various unfavorable environments.
Our work includes the collection of training and test data, decision on pre-processing of radar samples and the adaptation of a Convolutional Neural Network (CNN) to process radar data. Data acquisition and the adaptation of a CNN to time domain radar samples is our main focus, as CNNs are currently mainly used for image processing tasks. What distinguishes the radar samples from the image values is their complex value. Therefore, our RadarCNN or respectively its input must be adapted accordingly. 
The system aims to extend the possibilities of indoor perception and scene understanding with the information radar can provide.
Our contributions are as follows:
\begin{itemize}
    \item Indoor radar data collection for hand-sized objects
    \item Adaption of complex multi-channel time-domain radar data as CNN input
    \item Modified CNN layer structure suitable to process IQ radar data
\end{itemize}
Section \ref{relWork} deals with the related work in the field of indoor radar perception and imaging and addresses related automotive applications in terms of classification. Section \ref{meth} presents the process of data collection and our proposed classification method using radar data. Finally, the results, evaluation and discussion of the proposed methods are presented in Sections \ref{res} and \ref{dis}.

\section{Related Work}
\label{relWork}
Two main areas are considered in the current literature. Firstly, indoor perception using mmWave radar in a broader sense and secondly, detection and classification in automotive radar. Due to strong advances in autonomous driving, radar plays a crucial role in automotive applications. However, the main approach there is to use radar to detect and  classify dynamic objects in street scenarios \cite{autom1, autom2, autom3}, often using sensor fusion with cameras \cite{fusion1, fusion2}. However, the primary method employed here involves utilizing the radar's output signal to generate Range-Doppler or Range-Angle maps, which are then treated as images for the application of image processing and computer vision techniques.
Different representations of radar data are used for different classification and estimation tasks.
The automotive radar sector further spreads on more complex perception tasks like human activity understanding to further enhance road scene understanding \cite{human}.
\\
Indoor imaging radar applications of any kind are limited and there is few prior work on radar used for indoor perception. For example, Xu et al. \cite{Xu} and Drogu et al. \cite{Drogu} used mmWave radar for indoor mapping, while Montgomery et al. \cite{mont} and Khushaba et al. \cite{khus} used a radar sensor for the estimation of different surface materials.
\\
The indoor mapping mainly focuses on shape and depth retrieval of indoor environments while the material estimation relies on reflectivity patterns of different materials.
This draws the line to our small-scale object classifier. In order to be able to make an accurate distinction between detailed objects, sufficient information about the shape and also the surface and its reflectivity must be provided by the radar. This is necessary in order to be able to make a valid estimate across several object classes.
This can be achieved using a high-resolution 4D imaging radar, which is in our case the \textit{IMAGEVK-74} from \textit{Vayyar} and \textit{minicircuits.com} \cite{vayyar}.
The main reason to use this specific radar is the easy and handy way to collect time domain data as well as its high resolution in all degrees of freedom. This includes horizontal- and vertical angle resolutions of approximately $6.7 \degree$ and a best possible range resolution of $2.14$cm.
Comparable imaging solution from different manufacturers like Texas Instruments lack resolution especially in elevation angle compared to the \textit{IMAGEVK-74}. Mainly due to less Tx and Rx antennas, resulting in a lower number of virtual channels \cite{TI}.

\section{Proposed System and Methods}
\label{meth}

\subsection{Data Collection}
\begin{figure}
    \centering
    \fbox{\includegraphics[width=0.665\linewidth]{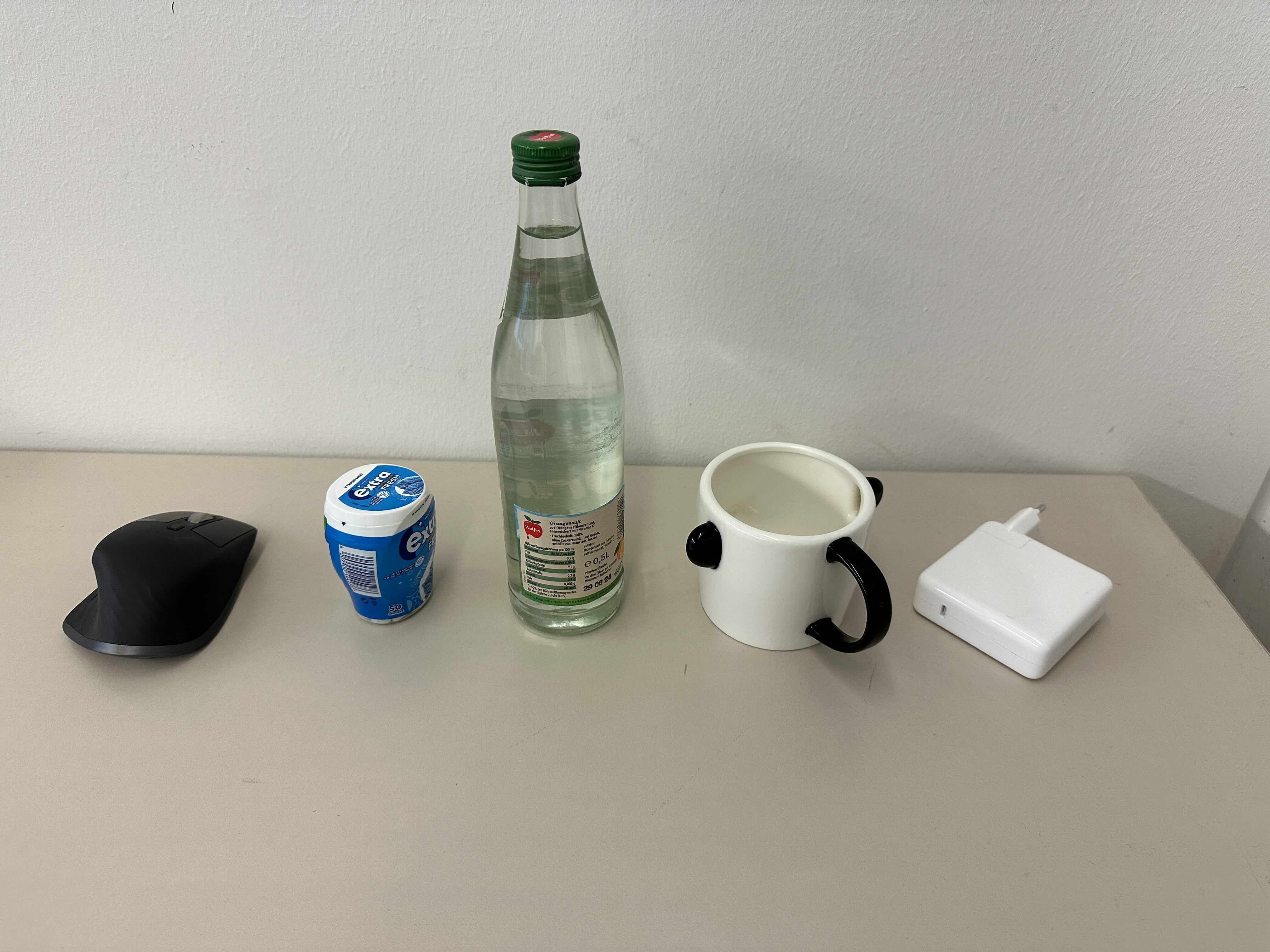}}
    \caption{Collection of the different objects.}
    \label{collection}
\end{figure}

The availability of public data sets in radar for indoor environments is very limited. It decreases even more when searching for data sets with raw ADC samples of radar scenes from high-resolution imaging radars.
\\
This created the need to record data and retrieve IQ samples from a radar sensor.
\begin{figure}
  \begin{subfigure}{0.225\textwidth}
    \fbox{\includegraphics[width=\linewidth]{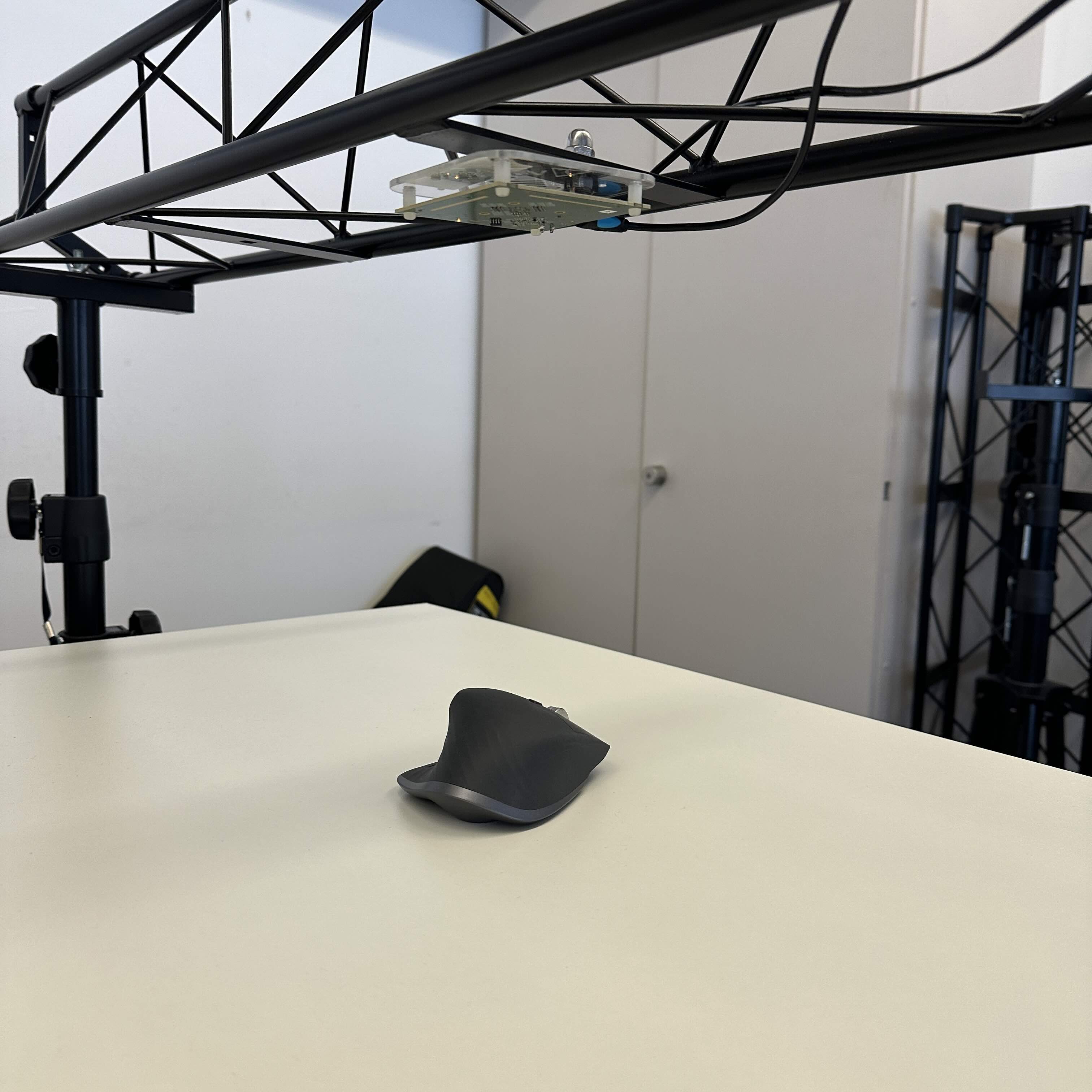}}
  \end{subfigure}%
  \hspace{0.02\textwidth}
  \begin{subfigure}{0.225\textwidth}
    \fbox{\includegraphics[width=\linewidth]{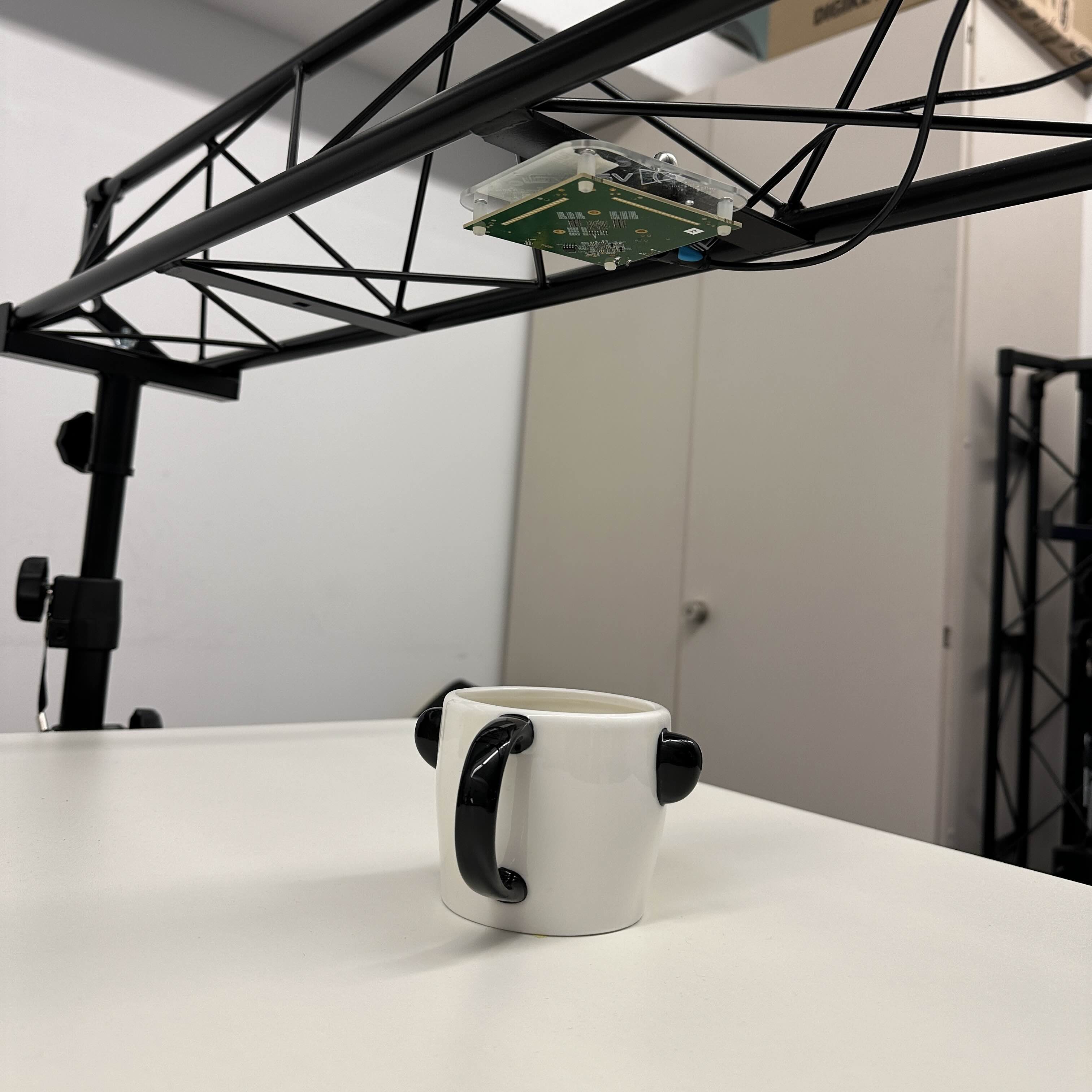}}
  \end{subfigure}
  \caption{Data collection process using different  objects in an indoor environment.}
  \label{testbed}
\end{figure}
Fig. \ref{testbed} depicts the testbed created and utilized for recording data.
\\
We recorded 400 radar frames for training and conventional testing with a single pulse per frame in total with $Tx \times Rx$ channels with $Tx=20$, $Rx=20$ and 100 samples per channel.
$Tx$ and $Rx$ represents the number of transmit and receive antennas of our radar, and a single reflected chirp is sampled with 100 points over time.
The resulting data set $\mathcal{X}$ consists of $n = 400$ elements with
\begin{equation}
    \mathcal{X} = \{x_{0}, x_{1},...,x_{n-1}\} \hspace{2cm} x_{i} \in \mathbb{C}^{20\times20\times100}
\end{equation}
The data set contains data of five different object classes with different shapes, material composition, and object poses. 
The five different objects are:
\begin{itemize}
    \item Ceramic cup
    \item Laptop charger
    \item Computer mouse
    \item Pack of chewing gum
    \item Water bottle
\end{itemize}
They are depicted in Fig. \ref{collection}.
The setup itself is static, which means every object stays in the same pose per single measurement frame.
Every frame captures the object placed in different orientation. Having five different object classes, there are 80 different images and object orientations per class.
The full data set splits into 300 training frames and 100 testing frames for the deep learning part.
The radar itself was configured with chirp bandwidth $B = 5$ GHz at center frequency $f_{c}=65.5$ GHz, and with sampling resolution of $f_{res}=10$ kHz.
Additionally, we recorded an another occluded testset in the same fashion. This was conducted with the same objects but with a carton box placed above the object. In this data set, an additional 200 samples are collected for testing only.
This data is aimed to use for cross domain testing of the system and is not used for training.
\\
Lastly, we augment the non-occluded testing data with additional white Gaussian noise to obtain the final testing set.
We use different noise levels in similar magnitude to the radar signal to evaluate noise resistance.
This serves also the purpose to test the system on impaired data.
In total, the data includes the original test set, a noisy version of the test set and an additional test set with concealed objects.

\subsection{Pre-processing and Neural Network Design}
\begin{figure}
    \centering
\begin{tikzpicture}
\pgfmathsetmacro{\cubex}{3}
\pgfmathsetmacro{\cubey}{3}
\pgfmathsetmacro{\cubez}{6}
\draw[black,fill=cyan] (0,0,0) -- ++(-\cubex,0,0) -- ++(0,-\cubey,0) --++(\cubex,0,0) -- cycle;
\draw[black,fill=cyan] (0,0,0) -- ++(0,0,-\cubez) -- ++(0,-\cubey,0) -- ++(0,0,\cubez) -- cycle;
\draw[black,fill=cyan] (0,0,0) -- ++(-\cubex,0,0) -- ++(0,0,-\cubez) -- ++(\cubex,0,0) -- cycle;

\draw [decorate, decoration={brace,amplitude=15pt,raise=1pt}] (-3,-3,0) -- ++(0,\cubey,0) node [midway, anchor= east, xshift=-4mm, outer sep=10pt]{Virtual Channels};

\draw [decorate, decoration={brace,amplitude=15pt,raise=1pt}] (-3,0,0) -- ++(0,0,-\cubez) node [midway, anchor= south east, outer sep=10pt]{Time Samples};

\draw[dashed] (-0.5,0,0) --++ (0,-3,0);
\draw[dashed] (-1,0,0) --++ (0,-3,0);
\draw[dashed] (-1.5,0,0) --++ (0,-3,0);
\draw[dashed] (-2,0,0) --++ (0,-3,0);
\draw[dashed] (-2.5,0,0) --++ (0,-3,0);

\draw[dashed] (0,-0.5,0) --++ (-3,0,0);
\draw[dashed] (0,-1,0) --++ (-3,0,0);
\draw[dashed] (0,-1.5,0) --++ (-3,0,0);
\draw[dashed] (0,-2,0) --++ (-3,0,0);
\draw[dashed] (0,-2.50,0) --++ (-3,0,0);

\draw[dashed] (0,-0.5,0) --++ (0,0,-6);
\draw[dashed] (0,-1,0) --++ (0,0,-6);
\draw[dashed] (0,-1.5,0) --++ (0,0,-6);
\draw[dashed] (0,-2,0) --++ (0,0,-6);
\draw[dashed] (0,-2.50,0) --++ (0,0,-6);

\draw[dashed] (-0.5,0,0) --++ (0,0,-6);
\draw[dashed] (-1,0,0) --++ (0,0,-6);
\draw[dashed] (-1.5,0,0) --++ (0,0,-6);
\draw[dashed] (-2,0,0) --++ (0,0,-6);
\draw[dashed] (-2.5,0,0) --++ (0,0,-6);

\draw[dashed] (0,0,-1) --++ (0,-3,0);
\draw[dashed] (0,0,-2) --++ (0,-3,0);
\draw[dashed] (0,0,-3) --++ (0,-3,0);
\draw[dashed] (0,0,-4) --++ (0,-3,0);
\draw[dashed] (0,0,-5) --++ (0,-3,0);

\draw[dashed] (0,0,-1) --++(-3,0,0);
\draw[dashed] (0,0,-2) --++(-3,0,0);
\draw[dashed] (0,0,-3) --++(-3,0,0);
\draw[dashed] (0,0,-4) --++(-3,0,0);
\draw[dashed] (0,0,-5) --++(-3,0,0);

\end{tikzpicture}
    \caption{Reference visualization of a $Rx\times Tx\times \#samples$ radar data cube with $Rx$-$Tx$ virtual channel alignment.}
    \label{3d_data_cube}
\end{figure}
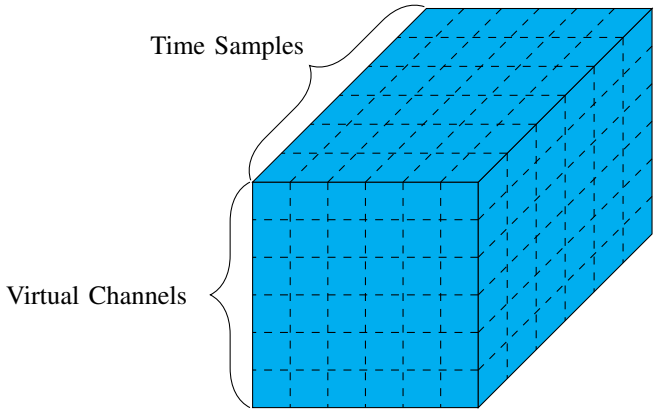

The starting point of the processing chain are frames from the radar sensor. The data is usually shaped as a radar data cube of $Tx$ antennas $\times$ $Rx$ antennas $\times$ number of samples. In total, there are 400 channels with 100 ADC samples per channel per frame. Such a data cube is depicted in Fig. \ref{3d_data_cube}.
Every frame contains information about distance, reflectivity power and also direction of arrival of the signal due to the antenna arrangement.
Since we capture only one pulse per frame, there is no information about dynamic movement of the scene. However, as we only work with static objects, this does not affect the system performance.
Instead of using conventional signal processing to extract information about Range-Angle or Cross-Angle, the samples are arranged in a $400\times100\times N$ pseudo data cube with depth $N$ and fed into a Convolutional Neural Network (CNN).
Pseudo here means that the data cube is not really a cube, but rather a $N=1$, or $N=2$ data layer.
Due to the CNNs feature extractions with subsequent convolutions, this network architecture is able to extract spatial features as well as temporal features from the raw radar signal.
\\
The neural network itself consists of three different 2D convolutional layers with a filter kernel size of $5\times5$.
These layers generate a multi-depth feature space from the input data tensor, depicted in Fig. \ref{flowchart}. Due to the rather small spatial size of the input data, a max pooling layer is only included after the last convolutional layer.
This is mainly because incorporating higher number of pooling layers results in an excessive reduction in the spatial dimension of the radar data, which is not needed here.
However, the final pooling layer still helps to increase robustness against sudden signal changes.
It also helps to optimize training and accentuate data features. After the last convolutional layer, the feature output is flattened and fed into a fully connected layer conducting the final decision.
The selected loss function for this classification task is the cross entropy loss, shown in (\ref{crossentropy}) along with the Adam optimizer and a learning rate of 0.001 \cite{adam}.
The number of classes $C$ is in our case $5$, and $\boldsymbol{\hat{y}}$ and $\boldsymbol{y}$ are the networks predicted distribution respectively the target distribution. 
\begin{equation}
    H(\boldsymbol{\hat{y}},\boldsymbol{y}) = - \sum_{j=1}^{C} y_{j} \log(\hat{y}_{j})
    \label{crossentropy}
\end{equation}
For training, we use batch size 16 including batch normalization after every layer to increase the learning capability of the network \cite{batch}. This improved the performance significantly.
Batch normalization in particular has helped our system to be able to learn in the first place, which makes this step a crucial point of the network design.
Finally, we use the ReLU function as activation function after every convolutional layer.
Since the radar samples consist of complex IQ values, the input for the network can either be the signal's real part, the imaginary part, or both. Overall, the full information content of the signal lies in both signal components combined. However, the single components also yield information about the captured scene.
\tikzstyle{input} = [rectangle,minimum width=2cm, minimum height=.5cm,text centered, draw=black]
\tikzstyle{layer} = [rectangle,minimum width=4cm, minimum height=.5cm,text centered, draw=black]
\tikzstyle{convlayer} = [rectangle,minimum width=5cm, minimum height=.5cm,text centered, draw=black]
\tikzstyle{arrow} = [thick,->,>=stealth]
\begin{figure}[ht]
    \centering

    \begin{tikzpicture}[node distance=.9cm]
    \node (data) [input] {Input Data};
    \node [below of=data, node distance=1.4cm] (batch1) [layer] {Batch Norm.};
    \node [below of=batch1] (layer1) [convlayer] {First Conv. Layer};
    \node [below of=layer1] (relu1) [layer] {ReLu Function};

    \node [below of=relu1, node distance=1.5cm] (batch2) [layer] {Batch Norm.};
    \node [below of=batch2] (layer2) [convlayer] {Second Conv. Layer};
    \node [below of=layer2] (relu2) [layer] {ReLu Function};

    \node [below of=relu2, node distance=1.4cm] (batch3) [layer] {Batch Norm.};
    \node [below of=batch3] (layer3) [convlayer] {Third Conv. Layer};
    \node [below of=layer3] (relu3) [layer] {ReLu Function};
    \node [below of=relu3] (maxpool) [layer] {2D Max Pool};

    \node [below of=maxpool, node distance=1.4cm] (flatten) [layer] {Flattening};
    
    \node [below of=flatten, node distance=1.4cm] (fc) [convlayer] {Fully Connected Layer};
    \node [below of=fc, node distance= 1.4cm] (output) [input] {Output};

    \draw [arrow] (data) --node[anchor=west] {\small$400\times100\times N$} (batch1);
    \draw [arrow] (batch1) -- (layer1);
    \draw [arrow] (layer1) -- (relu1);

    \draw [arrow] (relu1) --node[anchor=west] {\small$396\times96\times3$} (batch2);
    \draw [arrow] (batch2) -- (layer2);
    \draw [arrow] (layer2) -- (relu2);

    \draw [arrow] (relu2) --node[anchor=west] {\small$392\times92\times6$} (batch3);
    \draw [arrow] (batch3) -- (layer3);
    \draw [arrow] (layer3) -- (relu3);
    \draw [arrow] (relu3) -- (maxpool);

    \draw [arrow] (maxpool) --node[anchor=west] {\small$388\times88\times12$} (flatten);
    \draw [arrow] (flatten) --node[anchor=west] {\small$409728\times1$} (fc);
    \draw [arrow] (fc) --node[anchor=west] {\small$5\times1$} (output);
    
    \end{tikzpicture}
    \caption{Convolutional neural network architecture including data dimension of every step. $N=1$ for real or imaginary part only and $N=2$ for the use of both signal components.}
    \label{flowchart}
\end{figure}
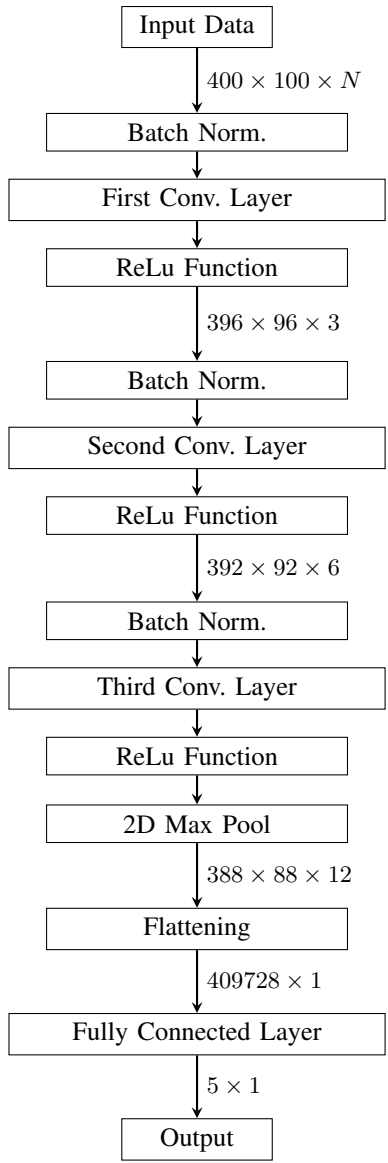

\section{Results}
\label{res}
The evaluation of our proposed RadarCNN is divided into three parts. First, the usual test part with 100 frames of new, unseen test data. We then test the system for its current robustness with regard to deviating data.
As mentioned, we evaluate over five different object classes, each of them with different shape and different material composition. All objects have approximately the same size.
Fig. \ref{realpart} shows the confusion matrix when training and testing the network using the real (In-Phase) part of the signal.
We managed to achieve a testset accuracy of 99\%. 
Similar results occur when using the imaginary (Quadrature) part.
Fig. \ref{imagpart} shows the respective results.
Combining both signal components is depicted in Fig. \ref{cplx}. The accuracy is similar to the single components, reaching 98\%.
\begin{figure}[ht]
    \centering
    \begin{tikzpicture}[scale=0.7]
    \begin{axis}[
            colormap={bluewhite}{color=(white) rgb255=(90,96,191)},
            xlabel=Prediction (in \%),
            xlabel style={yshift=-5pt},
            ylabel=Label,
            ylabel style={yshift=5pt},
            xticklabels={Cup, Charger, Mouse, Gum, Bottle}, %
            xtick={0,...,4}, %
            xtick style={draw=none},
            yticklabels={Cup, Charger, Mouse, Gum, Bottle}, %
            ytick={0,...,4}, %
            ytick style={draw=none},
            enlargelimits=false,
            xticklabel style={
              rotate=90
            },
            nodes near coords={\pgfmathprintnumber\pgfplotspointmeta},
            nodes near coords style={
                yshift=-7pt
            },
        ]
        \addplot[
            matrix plot,
            mesh/cols=5, %
            point meta=explicit,draw=gray
        ] table [meta=C] {
            x y C
            0 0 95
            1 0 0
            2 0 0
            3 0 5
            4 0 0
            
            0 1 0
            1 1 100
            2 1 0
            3 1 0
            4 1 0
            
            0 2 0
            1 2 0
            2 2 100
            3 2 0
            4 2 0
    
            0 3 0
            1 3 0
            2 3 0
            3 3 100
            4 3 0
    
            0 4 0
            1 4 0
            2 4 0
            3 4 0
            4 4 100
            
        }; %
    \end{axis}
\end{tikzpicture}
    \caption{Confusion Matrix with 99\% accuracy using the real part of the signal.}
    \label{realpart}
\end{figure}
\begin{figure}[ht]
    \centering
    \begin{tikzpicture}[scale=0.7]
    \begin{axis}[
            colormap={bluewhite}{color=(white) rgb255=(90,96,191)},
            xlabel=Prediction (in \%),
            xlabel style={yshift=-5pt},
            ylabel=Label,
            ylabel style={yshift=5pt},
            xticklabels={Cup, Charger, Mouse, Gum, Bottle}, %
            xtick={0,...,4}, %
            xtick style={draw=none},
            yticklabels={Cup, Charger, Mouse, Gum, Bottle}, %
            ytick={0,...,4}, %
            ytick style={draw=none},
            enlargelimits=false,
            xticklabel style={
              rotate=90
            },
            nodes near coords={\pgfmathprintnumber\pgfplotspointmeta},
            nodes near coords style={
                yshift=-7pt
            },
        ]
        \addplot[
            matrix plot,
            mesh/cols=5, %
            point meta=explicit,draw=gray
        ] table [meta=C] {
            x y C
            0 0 85
            1 0 0
            2 0 0
            3 0 15
            4 0 0
            
            0 1 0
            1 1 100
            2 1 0
            3 1 0
            4 1 0
            
            0 2 0
            1 2 0
            2 2 100
            3 2 0
            4 2 0
    
            0 3 0
            1 3 0
            2 3 0
            3 3 100
            4 3 0
    
            0 4 0
            1 4 0
            2 4 0
            3 4 0
            4 4 100
            
        }; %
    \end{axis}
\end{tikzpicture}
    \caption{Confusion Matrix with 97\% accuracy using the imaginary part of the signal.}
    \label{imagpart}
\end{figure}
\begin{figure}[ht]
    \centering
    \begin{tikzpicture}[scale=0.7]
    \begin{axis}[
            colormap={bluewhite}{color=(white) rgb255=(90,96,191)},
            xlabel=Prediction (in \%),
            xlabel style={yshift=-5pt},
            ylabel=Label,
            ylabel style={yshift=5pt},
            xticklabels={Cup, Charger, Mouse, Gum, Bottle}, %
            xtick={0,...,4}, %
            xtick style={draw=none},
            yticklabels={Cup, Charger, Mouse, Gum, Bottle}, %
            ytick={0,...,4}, %
            ytick style={draw=none},
            enlargelimits=false,
            xticklabel style={
              rotate=90
            },
            nodes near coords={\pgfmathprintnumber\pgfplotspointmeta},
            nodes near coords style={
                yshift=-7pt
            },
        ]
        \addplot[
            matrix plot,
            mesh/cols=5, %
            point meta=explicit,draw=gray
        ] table [meta=C] {
            x y C
            0 0 90
            1 0 0
            2 0 0
            3 0 10
            4 0 0
            
            0 1 0
            1 1 100
            2 1 0
            3 1 0
            4 1 0
            
            0 2 0
            1 2 0
            2 2 100
            3 2 0
            4 2 0
    
            0 3 0
            1 3 0
            2 3 0
            3 3 100
            4 3 0
    
            0 4 0
            1 4 0
            2 4 0
            3 4 0
            4 4 100
            
        }; %
    \end{axis}
\end{tikzpicture}
    \caption{Confusion Matrix with 98\% accuracy using both signal components.}
    \label{cplx}
\end{figure}
It can be observed that with all approaches, only the ceramic cup is more difficult to classify. All other objects can be estimated with 100\% accuracy.
Main factors for the high accuracy here are definitely the structural similarities of the different object samples in training and testing as well as the low amount of overall signal distortion.

\begin{figure}
    \centering
    \begin{tikzpicture}[scale=0.7]

\definecolor{darkgray176}{RGB}{176,176,176}
\definecolor{green01270}{RGB}{0,127,0}
\definecolor{darkturquoise0191191}{RGB}{0,191,191}
\definecolor{darkviolet1910191}{RGB}{191,0,191}
\definecolor{goldenrod1911910}{RGB}{191,191,0}

\begin{axis}[
tick align=outside,
tick pos=left,
x grid style={darkgray176},
xlabel={Noise Power $\sigma^2$},
xmajorgrids,
xmin=0, xmax=0.000025,
xtick style={color=black},
y grid style={darkgray176},
ylabel={Overall Accuracy ($\%$)},
ymajorgrids,
legend pos=south west,
ymin=20, ymax=100,
ytick style={color=black}
]
\addplot [thick, green01270]
table {%
0. 99
0.000001 99
0.000004 90
0.000009 60
0.000016 42
0.000025 35
};
\addlegendentry{Real Part}

\addplot [thick, darkviolet1910191]
table {%
0 97
0.000001 97
0.000004 96
0.000009 80
0.000016 71
0.000025 47
};
\addlegendentry{Imaginary Part}

\addplot [thick, red]
table {%
0 99
0.000001 99
0.000004 93
0.000009 50
0.000016 39
0.000025 34
};
\addlegendentry{Complex Signal}

\end{axis}

\end{tikzpicture}
    \caption{Overall accuracy when artificial Gaussian noise is added. The average signal power $\overline{P}_{s}$ is approx. $0.4\times10^{-3}$ for real and imaginary part.}
    \label{noise}
\end{figure}
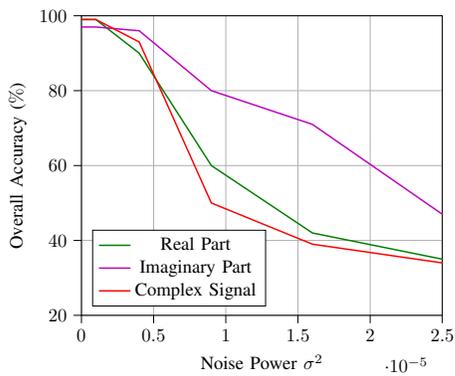

\begin{figure}
    \centering
    \fbox{\includegraphics[width=0.465\linewidth]{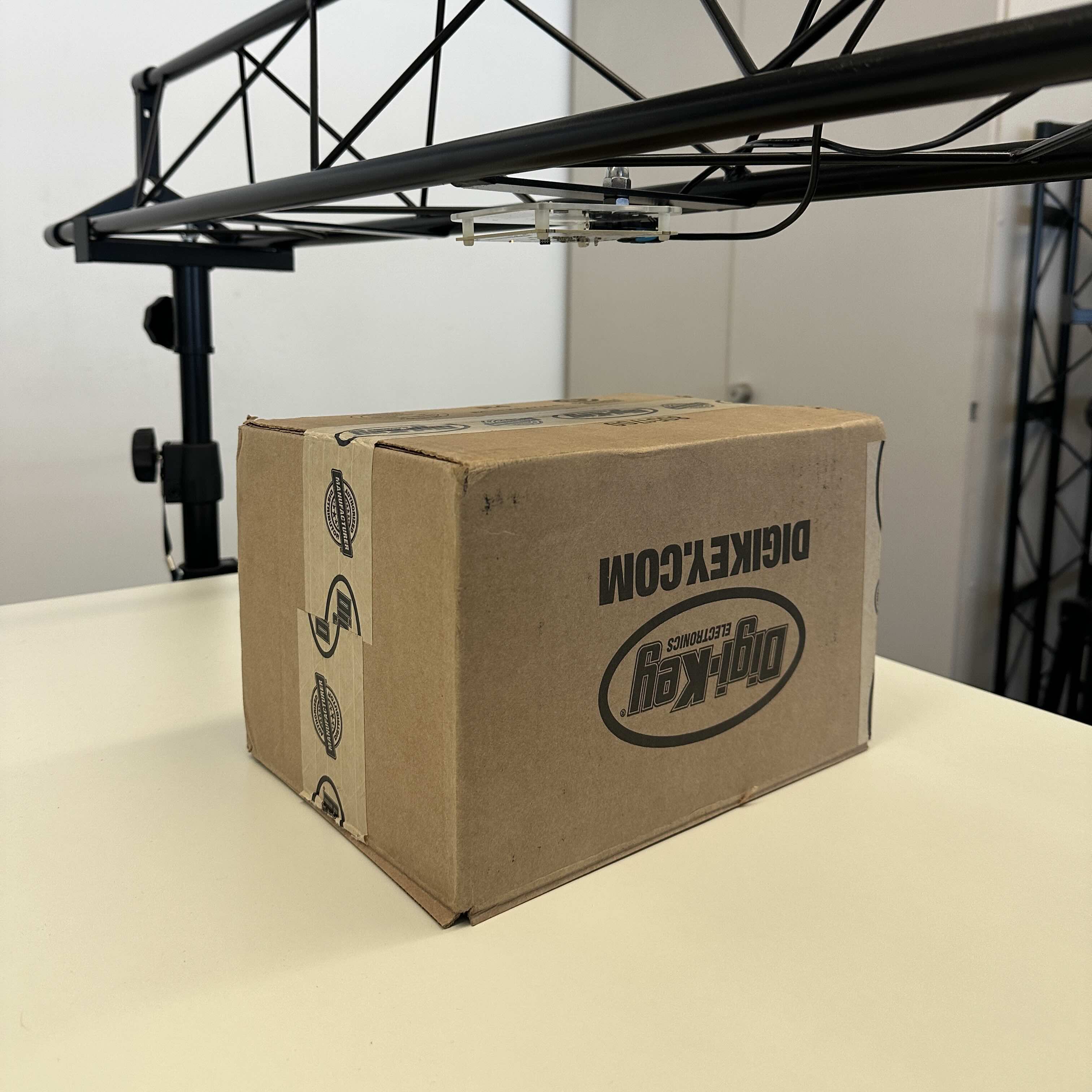}}
    \caption{Setup for data collection with occluded objects.}
    \label{testbed_occ}
\end{figure}

The second way to evaluate the system robustness is with addition of white Gaussian noise $\boldsymbol{Z}$ with variance $\sigma^2$ to the test data $\boldsymbol{X}$, without training the network again or modifying its parameters.
\begin{equation}
    \boldsymbol{Y} = \boldsymbol{X} + \boldsymbol{Z} \hspace{2cm}  \boldsymbol{Z}\sim \boldsymbol{\mathcal{C}} \boldsymbol{\mathcal{N}}(0,\,\sigma^{2})
\end{equation}
The noise is applied on the complex IQ samples $\boldsymbol{X}$.
Fig. \ref{noise} shows the overall accuracy with increasing noise power.
The accuracy remains high up to a certain point and then drops drastically. Up to this specific point, however, the system remains resilient.
This happens with each input signal variant.
Interesting is the difference in accuracy of the three different inputs. The Quadrature (imaginary) part of the signal seems to be more resilient than the other two configurations. However, this is probably due to our evaluation method, which selected the best possible system accuracy in terms of noise resilience.
An average accuracy profile would probably see the different input types coming closer together in terms of accuracy.
\\
The greatest impairment of the radar signal is the obscuring of the objects.
As depicted in Fig. \ref{testbed_occ}, the whole object is hidden under a carton box, while the radar sensor recorded.
The confusion matrices are depicted in Fig. \ref{conf_occ_real}, Fig. \ref{conf_occ_imag}, and Fig. \ref{conf_occ_cplx}.
\begin{figure}
    \centering
    \begin{tikzpicture}[scale=0.7]
    \begin{axis}[
            colormap={bluewhite}{color=(white) rgb255=(90,96,191)},
            xlabel=Prediction (in \%),
            xlabel style={yshift=-5pt},
            ylabel=Label,
            ylabel style={yshift=5pt},
            xticklabels={Cup, Charger, Mouse, Gum, Bottle}, %
            xtick={0,...,4}, %
            xtick style={draw=none},
            yticklabels={Cup, Charger, Mouse, Gum, Bottle}, %
            ytick={0,...,4}, %
            ytick style={draw=none},
            enlargelimits=false,
            xticklabel style={
              rotate=90
            },
            nodes near coords={\pgfmathprintnumber\pgfplotspointmeta},
            nodes near coords style={
                yshift=-7pt
            },
        ]
        \addplot[
            matrix plot,
            mesh/cols=5, %
            point meta=explicit,draw=gray
        ] table [meta=C] {
            x y C
            0 0 100
            1 0 0
            2 0 0
            3 0 0
            4 0 0
            
            0 1 5
            1 1 27.5
            2 1 0
            3 1 67.5
            4 1 0
            
            0 2 30
            1 2 40
            2 2 0
            3 2 30
            4 2 0
    
            0 3 0
            1 3 0
            2 3 0
            3 3 100
            4 3 0
    
            0 4 0
            1 4 100
            2 4 0
            3 4 0
            4 4 0
            
        }; %
    \end{axis}
\end{tikzpicture}
    \caption{Confusion Matrix using the real part of the IQ signal with 45.5\% accuracy with object occlusion.}
    \label{conf_occ_real}
\end{figure}
\begin{figure}
    \centering
    \begin{tikzpicture}[scale=0.7]
    \begin{axis}[
            colormap={bluewhite}{color=(white) rgb255=(90,96,191)},
            xlabel=Prediction (in \%),
            xlabel style={yshift=-5pt},
            ylabel=Label,
            ylabel style={yshift=5pt},
            xticklabels={Cup, Charger, Mouse, Gum, Bottle}, %
            xtick={0,...,4}, %
            xtick style={draw=none},
            yticklabels={Cup, Charger, Mouse, Gum, Bottle}, %
            ytick={0,...,4}, %
            ytick style={draw=none},
            enlargelimits=false,
            xticklabel style={
              rotate=90
            },
            nodes near coords={\pgfmathprintnumber\pgfplotspointmeta},
            nodes near coords style={
                yshift=-7pt
            },
        ]
        \addplot[
            matrix plot,
            mesh/cols=5, %
            point meta=explicit,draw=gray
        ] table [meta=C] {
            x y C
            0 0 100
            1 0 0
            2 0 0
            3 0 0
            4 0 0
            
            0 1 7.5
            1 1 0
            2 1 12.5
            3 1 80
            4 1 0
            
            0 2 0
            1 2 20
            2 2 32.5
            3 2 47.5
            4 2 0
    
            0 3 0
            1 3 0
            2 3 0
            3 3 100
            4 3 0
    
            0 4 2.5
            1 4 60
            2 4 15
            3 4 0
            4 4 22.5
            
        }; %
    \end{axis}
\end{tikzpicture}
    \caption{Confusion Matrix using the imaginary part the IQ signal with 51\% accuracy with object occlusion.}
    \label{conf_occ_imag}
\end{figure}
\begin{figure}
    \centering
    \begin{tikzpicture}[scale=0.7]
    \begin{axis}[
            colormap={bluewhite}{color=(white) rgb255=(90,96,191)},
            xlabel=Prediction (in \%),
            xlabel style={yshift=-5pt},
            ylabel=Label,
            ylabel style={yshift=5pt},
            xticklabels={Cup, Charger, Mouse, Gum, Bottle}, %
            xtick={0,...,4}, %
            xtick style={draw=none},
            yticklabels={Cup, Charger, Mouse, Gum, Bottle}, %
            ytick={0,...,4}, %
            ytick style={draw=none},
            enlargelimits=false,
            xticklabel style={
              rotate=90
            },
            nodes near coords={\pgfmathprintnumber\pgfplotspointmeta},
            nodes near coords style={
                yshift=-7pt
            },
        ]
        \addplot[
            matrix plot,
            mesh/cols=5, %
            point meta=explicit,draw=gray
        ] table [meta=C] {
            x y C
            0 0 97.5
            1 0 2.5
            2 0 0
            3 0 0
            4 0 0
            
            0 1 17.5
            1 1 80
            2 1 0
            3 1 2.5
            4 1 0
            
            0 2 0
            1 2 7.5
            2 2 0
            3 2 0
            4 2 92.5
    
            0 3 0
            1 3 7.5
            2 3 0
            3 3 92.5
            4 3 0
    
            0 4 0
            1 4 52.5
            2 4 0
            3 4 0
            4 4 47.5
            
        }; %
    \end{axis}
\end{tikzpicture}
    \caption{Confusion Matrix using both signal components with 63.5\% accuracy with object occlusion.}
    \label{conf_occ_cplx}
\end{figure}
Only the cup and the chewing gum can be classified correctly using both single signal components, as well as with a very high accuracy using the complex input. 
Some object classes cannot be classified correctly even once due to the large distortion.
The cup appears to be classified better compared with unimpared data, which is not straight forward explainable and probably highly dependent of the material used for concealment. The learned signal features from the cup seem to be stronger in the impaired version compared to the standard testset.
The gum is also detected with a high accuracy, which suggests its characteristic features still seem to be present with occlusion. All the rest is more or less heavily impaired.
However, considering that no adjustments were made to the training, the accuracy of 45-50\% exceeds random decision making by far, which shows part of the informational content still remaining in the measured radar signal.
This is also an advantage of mmWave radar over image sensors, as it is able to partially penetrate certain materials, such as cardboard in our example.

\section{Conclusion and Further Work}
\label{dis}
Our proposed radar classifier for small indoor objects and its cross domain performance shows how much information imaging radar can actually perceive in indoor environments. Especially the first evaluation with our default testset yields very good results and proofs radar as an indoor sensory device.
This is mainly because of the radar's imaging capability, which exceeds the performance of radars with fewer channels.
In addition, our method is also able to partially detect concealed objects, which is a major advantage over LiDAR or camera sensors. These are still more often the choice for indoor perception sensory.
However, resolution of the radar sensor and occurring distortion of the signal limits all radar applications, and also our classifier. This is even more significant when expanding such system to more capabilities like object shape retrieval.
The use of the full IQ signal should in theory provide  more information and thus higher accuracy of the environment, which can not finally be observed here.
However, sophisticated pre-processing of the IQ signal and prior extraction and refinement of its provided information should enhance the performance of the complex signal as input significantly.
\\
At the end, it is important to deploy imaging radars on suitable tasks in indoor environments in such a way that its possible impairments have as little influence as possible on the result.
This can open a lot new opportunities using radar sensory.

\end{document}